\documentclass[runningheads]{llncs}

\usepackage[T1]{fontenc}
\usepackage[utf8]{inputenc}
\usepackage{graphicx}
\usepackage{booktabs}
\usepackage{multirow}
\usepackage{tabularx}
\usepackage{array}
\usepackage[table]{xcolor}
\usepackage{float}
\usepackage{placeins}
\usepackage{subcaption}
\usepackage{calc}
\usepackage{listings}
\usepackage{pgfplots}
\usepackage{tikz}
\usetikzlibrary{positioning, shapes.geometric, arrows.meta, trees}
\usepackage[linesnumbered,ruled,vlined]{algorithm2e}
\usepackage{amsmath}

\usepackage{amssymb}
\usepackage{amsfonts}
\usepackage{amstext}
\usepackage{url}
\usepackage{orcidlink}
\renewcommand{\orcidID}[1]{\,\orcidlink{#1}}

\pgfplotsset{compat=1.18}

\begin{document}

\title{Moving Target Defense in SDN-enabled EV Charging Networks}
\titlerunning{Moving Target Defense in SDN-enabled EV Charging Networks}
\author{Roland Plaka\orcidID{0009-0000-5847-6317} \and
Mikael Asplund\orcidID{0000-0003-1916-3398} \and
Simin Nadjm-Tehrani\orcidID{0000-0002-1485-0802}}
\authorrunning{R. Plaka et al.}
\institute{Department of Computer and Information Science,\\
Linköping University, Sweden\\
\email{\{roland.plaka,mikael.asplund,simin.nadjm-tehrani\}@liu.se}}
\maketitle

{\renewcommand{\thefootnote}{}%
\footnotetext{This work was partially supported by the Swedish Energy Agency
through the RECAP project (P2025-01312) and by the Swedish Foundation for
Strategic Research (SSF), project FUS21-0033 (ASTECC).}}

\begin{abstract}
Software-Defined Networking (SDN) is emerging as a promising technology for EV Charging infrastructure (EVCI) since it enables flexible, programmable
control of EVCI networks, yet its security gain to EVCI remains underexplored. SDN-enabled EVCI faces an
increasing number of cyber threats arising from the convergence of
IT and OT components; in particular, low-rate Denial-of-Service
(DoS) attacks. Earlier works identify a trend toward low-rate attacks in OT networks that may exhaust SDN switch flow tables and which can result in 
disabling the entire charging sites without triggering volume-based
defenses.
In this paper, we propose CS-SHIELD, 
a Moving Target Defense mechanism for SDN-enabled EVCI communication. We showcase the implementation of the proposed mechanism that detects malicious flow table rules via cross-layer identity verification and responds by
reassigning virtual IP addresses to all active chargers.
Experiments on an emulated SDN testbed with a real charging protocol
implementation demonstrate that CS-SHIELD maintains full site
availability under attack, responds quickly, adds
only a negligible delay under normal conditions.
The reassigning (shuffle) approach at each reaction makes the earlier reconnaissance knowledge by the attacker rendered worthless.

\keywords{moving target defense \and DoS attack \and SDN security \and EV charging}
\end{abstract}

\sloppy
\setlength{\textfloatsep}{4pt plus 1pt minus 2pt}
\setlength{\floatsep}{3pt plus 1pt minus 1pt}
\setlength{\intextsep}{4pt plus 1pt minus 2pt}
\setlength{\abovecaptionskip}{3pt}
\setlength{\belowcaptionskip}{1pt}

\section{Introduction}
Electric Vehicle Charging Infrastructure (EVCI) is regarded
as critical infrastructure ~\cite{sarieddine2023ev,szakaly2025current}, and like other critical infrastructure it
can be a target for attack. It is a distributed cyber-physical system of
converged IT and OT components, in many cases accessible over the public
internet, with security weaknesses that have been reported in prior
studies~\cite{szakaly2025current}. The main concern in this domain is not the confidentiality of its
data but the consequence of disrupting it: a disturbance at the network
layer can translate into a physical loss of charging service. In such
cases an adversary need not break confidentiality or forge messages;
denying the control traffic can be enough to prevent vehicles from
charging.

Classical denial-of-service attacks exhaust resources with high-volume traffic and are detected by 
throughput and packet-rate monitors. EVCI, by contrast, runs low-bandwidth, periodic traffic over
long-lived TLS/WebSocket sessions governed by the Open Charge Point
Protocol (OCPP), the application-layer protocol that manages
communication between each charging station (CS) and the Charging Station
Management System (CSMS). Each charger maintains a persistent OCPP
session, sending periodic Heartbeat messages to signal availability;
a charger that misses its Heartbeat window is marked offline by the
CSMS. Because availability depends on the continuity of these
sessions rather than on raw bandwidth, an attacker may deny service
at a low rate without producing volume that conventional defences
would flag.
SDN is well suited for managing charging sites at scale: a single
programmable controller configures all switches centrally and
enforces consistent traffic policy across the site. However, an
SDN switch forwards traffic using a flow table of fixed capacity.
This table can be exhausted by a low-rate stream of crafted flows
rather than by high-volume traffic. EVCI over SDN can in some cases be problematic because of SDN being susceptible to low-rate DoS and the consequences of EVCI are very direct: an attacker who slowly
fills the flow table of a single switch could disable a charging site
without directly touching any charger or generating detectable
volume. 

Defending such situations is difficult because several common
mechanisms are a poor fit for an attack that operates below volume thresholds and 
targets long-lived sessions. Rate limiting and filtering may block
legitimate charger traffic and cause a self-inflicted outage;
volume-based detection can miss low-rate interference; static firewalls
may not adapt; and dynamic defences designed for short-lived
connections can disrupt the long-lived sessions EVCI relies on. Prior
work points to this gap. Sarieddine et al.~\cite{sarieddine2023ev} note
a lack of network-layer defences against remote intrusion, alongside
weaknesses in security procedures among manufacturers and operators.
Szakály et al.~\cite{szakaly2025current} report similar findings in a
large-scale measurement of deployed charging stations, 
Vailoces et
al.~\cite{vailoces2023securing} propose interface countermeasures,
though without detailed implementation guidance and requiring
coordination across operators. Many existing defences are applied
locally to individual components, which points to a need for a more holistic network-level approach. The problem we address is how to
preserve session continuity under low-rate network interference, without
relying on volume thresholds and without disrupting communication. We expand the notion of session continuity in OCPP ~\cite{iec63584_2024} to include  that each IP in the uniquely generated sequence, maps to the same session identifier over that session’s lifetime. In particular, for any active session, there always exists a unique mapping from a live IP to a session identifier.

Earlier work~\cite{mudgal2024flora} describes the flow-table-exhaustion
attack as a two-phase process: inferring the switch idle timeout, then
filling the table with long-lived flows. We extend this attack with a reconnaissance phase that
precedes both, in which the attacker first need to discover the targets. This attacker prerequisite is relevant to our
defence, since the address bindings the attacker learns during
reconnaissance are what the defence aims to invalidate at each
reconfiguration.

Moving Target Defense (MTD) directly counters this problem by
shifting system configurations over time, reducing the
predictability that the attack relies on~\cite{jajodia2011moving}. More recent work
illustrates this principle in practice: Pagnotta et al.~\cite{pagnotta2023dolos}
propose DOLOS, an MTD architecture that combines reconfiguration with
cyber deception and intrusion detection to mislead and disrupt an
attacker's reconnaissance. A challenge
when applying MTD to critical infrastructure is that the defensive
reconfiguration should not itself compromise availability, which is the
property under attack. Our research question is how to apply MTD while
keeping legitimate services available under a low-rate DoS attack. To the
best of our knowledge, MTD has not previously been studied for EV
charging infrastructure for this purpose. This paper
makes the following contributions:

\begin{enumerate}
    \item CS-SHIELD: an SDN-enabled mechanism within a  novel MTD framework to mitigate low-rate DoS attacks
    against EV charging infrastructure.
    \item A testbed including the SDN-enabled EVCI system, used to
    validate the attack model and demonstrate that flow table
    saturation causes measurable degradation of charging service
    availability.
    \item An empirical evaluation demonstrating that CS-SHIELD
    maintains charging availability under attack while imposing
    negligible overhead under normal operation. In particular, first, we evaluate to what extent low-rate DoS attacks can impact charging availability, and to what extent CS-SHIELD mitigates such attacks, while investigating the negative impact and overhead introduces to normal charging operation.
\end{enumerate}

In the remainder of the paper, we review and discuss the related work looking at the effectiveness and applicability
of MTD in different domains. Section 3 present the system and threat model. Section 4 describes the proposed
SDN-enabled MTD framework. Section 5 outlines the testbed implementation and attack method. Section 6, evaluates the attack and CS-SHIELD through an emulated use case. Section 7 serves as a conclusion to this paper.

\section{Related Works}

MTD works by changing system configurations (e.g., IP addresses,
ports, services, OS, network topologies) to confuse attackers,
exhaust their resources, and reduce the attack surface, addressing
\textit{what} to change, \textit{how}, and \textit{when}. Sun et
al.~\cite{sun2023survey} survey MTD techniques and categorize them
by affordability, optimization, and adaptability, providing a
taxonomy that contextualizes our approach.

\textbf{Address Space Randomization}
Yoon et al.~\cite{yoon2020attack} shuffle host
configurations by criticality in SDN to protect \emph{detection} of
critical hosts, whereas we shuffle to preserve \emph{availability}.
Potteiger et al.~\cite{potteiger2022tutorial} survey MTD for
automotive CPS and mention address-space randomization only
conceptually.
Reti et al.~\cite{reti2022evaluating} quantitatively
evaluate honeypot and MTD effectiveness in simulation, guiding
deployment and mutation-interval choices with low overhead, but for
simulated honeypots rather than EVCI-like components.
Burow et al.~\cite{burow2020moving} evaluate worst-case performance
of address-space randomization in real-time systems, finding
significant tail latencies that may matter for EVCI given its
connection to the power grid, though they do not consider an
attacker model. 

\textbf{Diversity}
Navas
et al.~\cite{navas2020mtd} survey MTD for IoT and conclude the field
is feasible but immature, without targeting a specific attack. Rehman et al.~\cite{rehman2024proactive} combine decoy nodes and OS
diversity so the attacker's view is continually misled by fake
targets; we instead randomize the addresses of the \emph{real}
chargers, invalidating reconnaissance already performed against
legitimate targets. Ma et al.~\cite{ma2022game} propose a
deception attack surface metric and game model; being applied and
domain-specific, our work is orthogonal to such theoretical
approaches. 

\textbf{Route-based}
Cyber deception uses crafted misinformation to mislead attackers,
shifting the defender--attacker asymmetry by increasing the time and
effort an attack requires. 
Abdelkhalek et al.~\cite{abdelkhalek2022moving} mitigate DoS via path
switching in smart-grid SDN, reducing packet drop rate; we instead
target flow-table exhaustion, detected via cross-layer identity
verification against OCPP's own session state, and measure
availability and forwarding delay since OCPP's persistent WebSocket
sessions must survive reconfiguration without being torn down. 
Etxezarreta et al.~\cite{etxezarreta2026replica} extend
SDN-assisted MTD to defend industrial control systems against False
Data Injection via proactive packet cloning and adaptive path
routing, introducing defensive uncertainty without degrading network
performance. 

\textbf{Other approaches}
Sharma~\cite{sharma2025evaluating} evaluates shuffling and
diversity-based MTD in IoT via time-to-compromise, showing that more
frequent shuffling increases attacker effort. Their periodic,
time-based shuffling suits smart-home settings where fixed-interval
reconfiguration does not disrupt ongoing sessions; in EVCI it would
interrupt active OCPP sessions at every interval, the same effect the
attack produces, which is why a reactive trigger like CS-SHIELD's is
necessary. 
Masud et al.~\cite{masud2025vulnerability} combine three
MTD techniques against IoT reconnaissance, with outcomes of either
releasing the asset to the attacker or interrupting it; releasing a
non-critical IoT device is tolerable, but releasing a charging point
is not; given EVCI's asymmetry and multi-stakeholder, grid-connected
nature, we adapt their security metrics to EVCI instead.
Assen et al.~\cite{assen2023lightweight} propose a
Falco-based MTD for multi-purpose IoT malware, monitoring system
calls for anomalous behaviour;
Soussi et
al.~\cite{soussi2023merlins} present deep-RL-enhanced MTD for NFV in
SDN. Both confirm MTD's feasibility in constrained settings,
supporting our choice of a lightweight reactive shuffle for EVCI.

Table~\ref{tab:related_work} positions CS-SHIELD against these works
along four dimensions. \textbf{SDN-type}: proactive (rules installed
before an attack), reactive (installed in response to an event), or
absent. \textbf{Domain}: IoT, Smart Grid, ICS, NFV, general IT,
Automotive, or EVCI. \textbf{Attack}: reconnaissance, DoS,
topology-based, malware, or FDI (\emph{unspecified} where no
concrete attack is targeted). \textbf{Solution}: Address Space
Randomisation (ASR), route switching, OS-diversity, SW-diversity,
path routing, or Deep-RL MTD. Our work is the only entry combining a
reactive SDN-type with an EVCI domain and a concrete low-rate DoS
threat addressed through ASR.

\begin{table}[htbp]
\centering
\caption{MTD approaches for SDN-enabled systems, grouped by solution}
\label{tab:related_work}
\scriptsize
\setlength{\tabcolsep}{3pt}
\renewcommand{\arraystretch}{0.95}
\scalebox{0.95}{
\begin{tabular}{@{}l l l l l@{}}
\toprule
\textbf{Solution} & \textbf{Author} & \textbf{SDN} & \textbf{Domain} & \textbf{Attack} \\
\midrule
\multirow{4}{*}{ASR}
  & Yoon~\cite{yoon2020attack}             & proactive & IT            & recon \\
  & Reti~\cite{reti2022evaluating}         & no SDN    & NA            & recon \\
  & Potteiger~\cite{potteiger2022tutorial} & no SDN  & Automotive    & unspecified \\
  & Burow~\cite{burow2020moving}           & no SDN    & Real-time CPS & unspecified \\
\cmidrule(l){1-5}
\multirow{2}{*}{SW diversity}
  & Navas~\cite{navas2020mtd}              & reactive  & IoT           & unspecified \\
  & Ma~\cite{ma2022game}                   & no SDN    & NA            & recon \\
OS diversity
  & Rehman~\cite{rehman2024proactive}      & proactive & IoT           & recon \\
\cmidrule(l){1-5}
Route switching
  & Abdelkhalek~\cite{abdelkhalek2022moving} & proactive & Smart Grid  & DoS \\
Path routing
  & Etxezarreta~\cite{etxezarreta2026replica} & proactive & ICS        & Data inj. \\
\cmidrule(l){1-5}
\multirow{3}{*}{Hybrid}
  & Masud~\cite{masud2025vulnerability}    & proactive & IoT           & recon \\
  & Assen~\cite{assen2023lightweight}      & reactive  & IoT           & malware \\
  & Sharma~\cite{sharma2025evaluating}     & no SDN    & IoT           & unspecified \\
Deep-RL
  & Soussi~\cite{soussi2023merlins}        & reactive  & SDN       & memory corr. \\
\midrule
\textbf{ASR} & \textbf{Ours} & \textbf{reactive} & \textbf{EVCI} & \textbf{recon, low-rate DoS} \\
\bottomrule
\end{tabular}}
\end{table}

\section{SDN-enabled EVCI}
This section describes a generic SDN-enabled EV charging infrastructure: its elements and their roles, the rules governing system behaviour, and a threat model covering attacker capabilities, goals, and workflow.
\subsection{System Model}
\label{sec:cps}

A cyber-physical system (CPS) closely couples computing and
communication to a physical process: the cyber part senses the
physical state, decides, and acts on the physical world, and that
state is fed back to the cyber part~\cite{rajkumar2010cps}. EV
charging infrastructure has this structure: charging is physical,
but no energy is delivered without communication, since a session is
authorized, sustained, and bounded by control messages. A fault in
the cyber part therefore reaches the physical part quickly: a lost
or delayed message can end a session or mark a charger offline,
making availability one of the main concerns.

\begin{figure}[htbp]
  \centering
  \begin{subfigure}[b]{0.47\columnwidth}
    \centering
    \includegraphics[width=\linewidth]{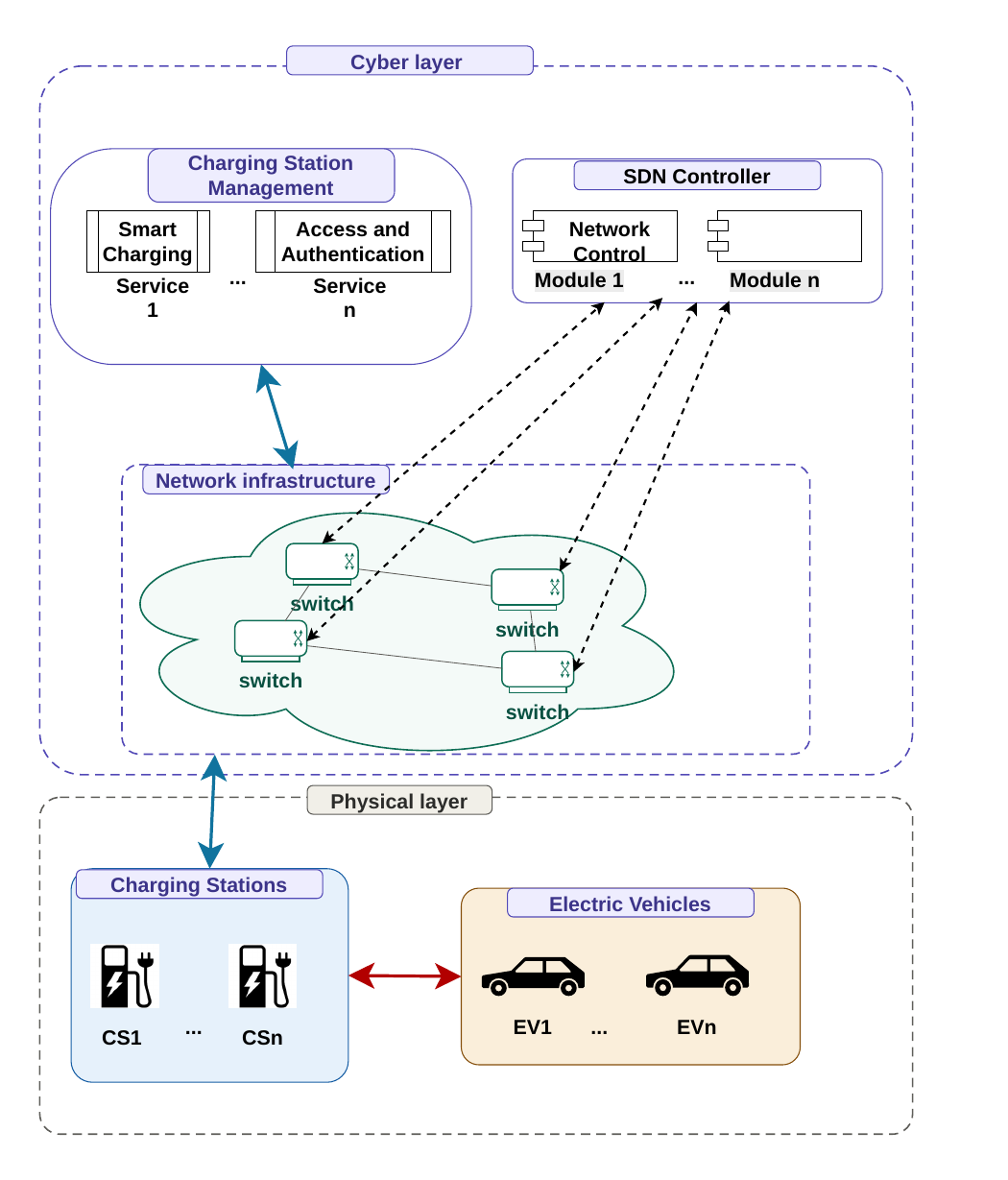}
    \caption{Cyber-physical model.}
    \label{fig:cps}
  \end{subfigure}\hfill
  \begin{subfigure}[b]{0.47\columnwidth}
    \centering
    \includegraphics[width=\linewidth]{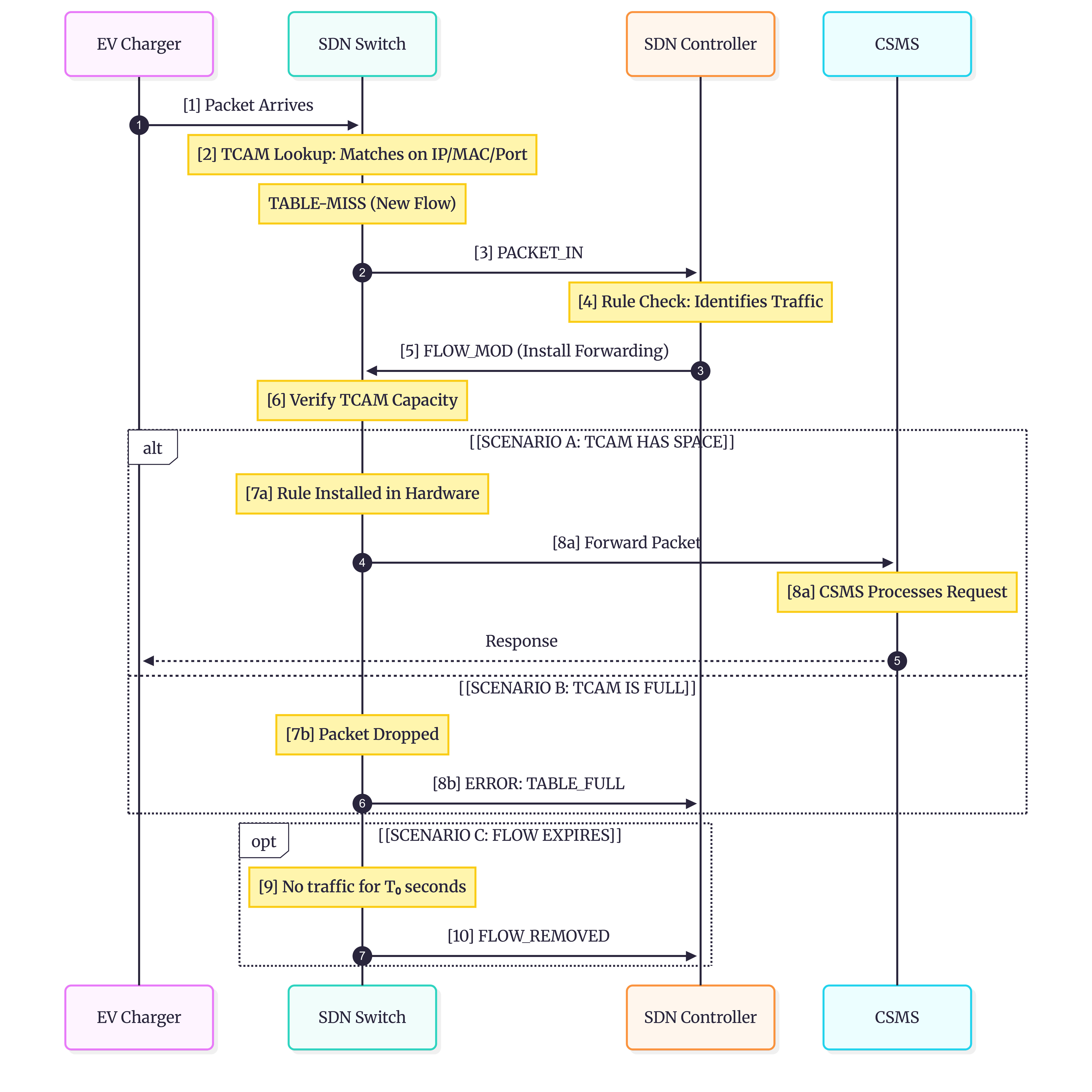}
    \caption{Reactive OpenFlow operation.}
    \label{fig:Normal SDN operation}
  \end{subfigure}
  \caption{SDN-enabled EVCI model and reactive forwarding behaviour.}
  \label{fig:cps_openflow}
\end{figure}

We model the infrastructure as two layers, shown in
Figure~\ref{fig:cps}. The \textbf{physical layer} hosts the charging
stations, which deliver energy and report status, and the electric
vehicles they serve; energy flows between the two during a session
(the red link), and the state of this layer is the ground truth the
rest of the system manages. The \textbf{cyber layer} observes and governs the physical layer and
is structured by Software-Defined Networking (SDN), which suits a
charging site that is large, distributed, and changing: a single
controller can program every switch from one point instead of
configuring each by hand. The cyber layer has two parts. The network
infrastructure is a set of SDN switches that forward packets purely
according to controller-installed rules, forming the path every
control message must cross. Above them, the SDN controller, the
network brain, is composed of independent modules for rule
installation and path selection, flow-statistics monitoring,
topology discovery, and access control, and can map virtual to real
addresses. The Charging Station Management System (CSMS) is the
charging manager: it runs charging services, monitors station
status, schedules sessions, and issues start/stop commands. The
controller programs switches over OpenFlow, while the CSMS talks to
chargers over OCPP; both are shown as blue links in the figure,
contrasting with the physical power link (red).

Figure~\ref{fig:Normal SDN operation} illustrates normal EVCS--CSMS
communication on a reactive SDN, where flow rules are installed on demand.
When a charger sends a packet, the switch checks for a matching flow rule.
If none exists, the switch forwards the packet to the controller, which
evaluates it against its policy. Three outcomes are possible. In
Scenario~A the flow table has space, so the controller installs the rule
and the packet is forwarded normally. In Scenario~B the flow table is
full, so the switch can install no further rules, drops the packet, and
the controller receives a table-full notification. In Scenario~C a rule
that has already been installed receives no matching traffic within the
idle timeout $T_o$, which is typically set to the OCPP Heartbeat interval of 60\,s,
so the rule expires, the controller is notified, and the charger must
re-establish its flow.
Two of these outcomes matter for the rest of the paper. The bounded flow
table of Scenario~B means a switch can be driven to a state where it
serves no new charger flows, and the timeout behaviour of Scenario~C
means each charger depends on its Heartbeat traffic to keep its rule
installed. In other words, charging availability depends on the flow
table capacity and charger flow rules.

\subsection{Threat Model}
\label{sec:threat}

The adversary aims to deny charging availability at a site by exhausting
the flow table of an SDN switch, so that the switch can no longer install
the forwarding rules that legitimate OCPP traffic requires. This follows
the Slow-TCAM attack of Pascoal et al.~\cite{pascoal2020slow}, which
exhausts a switch's Ternary Content Addressable Memory (TCAM) with low-rate traffic rather than high-volume
floods. The adversary is positioned on the network, co-located with the chargers,
for example by joining a site's shared WiFi. They
hold a valid IP address, can send and receive arbitrary TCP/IP packets,
can scan the network to discover hosts and open ports, and can open and
maintain many concurrent connections to a charger, all at a low packet
rate. The access point only exposes an SSID that follows the WPA3 Enterprise security scheme. As a result, the adversary can passively listen to and collect network frames, but cannot identify the IP addresses in a frame, except for their own~\cite{10274082}. In other words, the adversary cannot modify or read flow rules, impersonate a charger or the
CSMS, break TLS, passively infer the IP addresses of other nodes in the network or communicate with the controller. We assume the SDN
controller, the control channel, and the CSMS are trusted and not
compromised.

Each charger requires at least one flow-table rule to forward its
OCPP traffic. With a flow table of capacity $N_r$, the adversary
fills $N_r - N_c$ entries with crafted flows, leaving fewer than
$N_c$ slots for legitimate charger rules. Once the table is full,
new charger flows are dropped and a charger that misses its
Heartbeat window is marked offline by the CSMS. The attack also depends on knowledge, not only capability: to fill the
table against a charger, the adversary must first learn which hosts are
chargers and at which addresses they answer. We model the acquisition of
this charger-to-address binding as an explicit part of the adversary's
required knowledge, in contrast to prior work that assumes the targets are
known. Network-layer exhaustion of the flow table is our focus;
application-layer weaknesses in the OCPP software are left out of
scope~\cite{boussaha2025emuocpp}. The concrete attack procedure and
parameters appear with the evaluation in Section~\ref{sec:eval}. By cross-checking the switch's flow table against the chargers the CSMS has
authenticated, the controller identifies any address that
is carrying traffic yet was never authenticated as a charger.

\section{CS-SHIELD scheme}
\label{sec:mtd}

\begin{figure}[htbp]
    \centering
    \includegraphics[width=0.5\columnwidth]{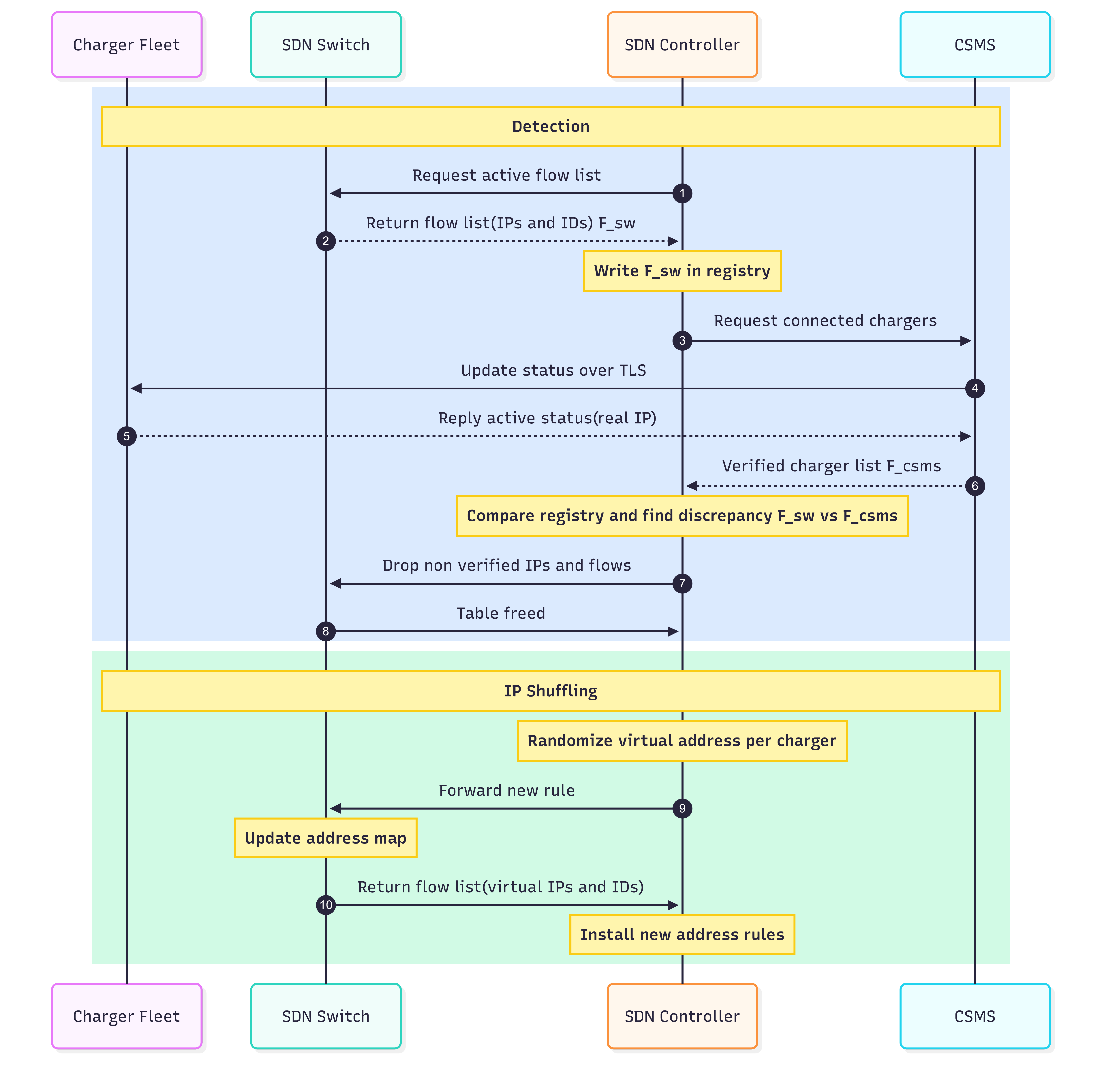}
    \caption{CS-SHIELD overview: detection (phase 1) and shuffling
    (phase 2)}
    \label{fig:mtd_sequence}
\end{figure}

An overview of our proposed CS-SHIELD scheme is presented in
Figure~\ref{fig:mtd_sequence}. CS-SHIELD (Charging Station
Software-defined Host IP Evasion for Low-rate Defense) is an MTD
mechanism that defends SDN-enabled EVCI against the flow-table exhaustion
attack of Section~\ref{sec:threat}. The scheme is structured into two
phases: (1) detection and (2) shuffling. The primary entities
participating in the scheme are the SDN switches, the controller, and the
CSMS and the chargers. During the scheme, mutual authentication is
established among the participating entities using TLS, as supported by
the OCPP and OpenFlow protocols. Notably, TLS communication is not
established between the attacker and the CSMS at the OCPP layer, and this
gap is what the detection phase exploits. CS-SHIELD runs on the SDN
controller and builds on the reactive operation of
Figure~\ref{fig:Normal SDN operation}, which it extends rather than
replaces. Shuffling is required so that the knowledge the attacker has obtained from previous attempts cannot be reused in subsequent attacks. In fact, even an attacker the connection of which has been dropped can initiate a new connection and reuse the information gained from the initial attack since previous information remains actionable.
The two phases are described below.

{\textbf{Phase 1: Detection}}
The detection phase of Figure~\ref{fig:mtd_sequence} identifies 
attacker-injected flows. The controller polls the switch for 
its active flows at a fixed interval $T_{\mathit{poll}}$, 
regardless of attack activity (step~1); the defense itself is 
triggered only once this polling reveals that the attacker's 
injected flows have filled the switch's flow table. The switch 
returns the full list of active flows, identified by IP address 
and flow ID, as $F_{sw}$ (step~2), which the controller writes 
to an internal registry (step~2, Write $F_{sw}$ in registry). 
The controller then requests the list of authenticated chargers 
from the CSMS (step~3). The CSMS asks each charger to report 
its real address over its authenticated OCPP channel (step~4), 
and each charger replies with its active status and real IP 
(step~5). The CSMS returns the verified charger list $F_{csms}$ 
to the controller (step~6). Because the attacker can send 
traffic but cannot authenticate to the CSMS, it appears in 
$F_{sw}$ but not in $F_{csms}$. The controller cross-checks 
the two: any address present in $F_{sw}$ but absent from 
$F_{csms}$ is flagged as malicious (step~6, Compare registry 
and find discrepancy). The controller then drops all 
non-verified flows from the switch (step~7), which frees the 
flow table (step~8). This cross-plane comparison distinguishes 
an attacker-injected flow from a legitimate charger flow 
without payload inspection.


{\textbf{Phase 2: Shuffling}}
In the second phase, the controller acts on the freed table to 
reassign the network addresses of all active chargers. For each 
charger, a new virtual IP address is drawn at random from a 
large address pool (step~8, Randomize virtual address per 
charger). The controller forwards the new forwarding rule to 
the switch (step~9) and updates its internal address map 
(step~9, Update address map). The switch returns the updated 
flow list with the new virtual IPs and flow IDs (step~10), 
and the controller installs the new address rules for all 
chargers (step~10, Install new address rules). An attacker 
who has discovered a charger's address during reconnaissance 
can no longer use it and must scan the network again. The 
change is unpredictable because the new address is random, 
and it is bounded because it touches only addressing and only 
at the moment of attack, which keeps it within EVCI's 
availability and latency constraints. Like Phase~1, this phase 
is triggered reactively by the detected discrepancy rather 
than on a fixed schedule.

\section{Testbed and Attack Implementation}
\label{sec:setup}

This section describes the testbed used to evaluate CS-SHIELD
against the Slow TCAM Exhaustion attack: reproducing the three-phase
attack in a controlled environment and measuring its impact on
availability, flow-table rule count, and forwarding delay. Symbol
definitions are given in Appendix~\ref{app:tables}.

\subsection{The experiment testbed}
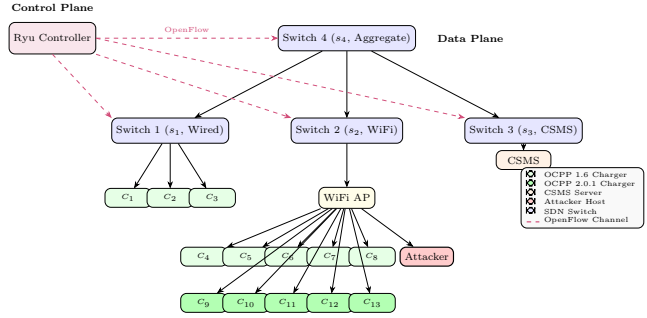
\begin{figure}[H]
\centering
\resizebox{0.70\columnwidth}{!}{
\begin{tikzpicture}[
    node distance=1.0cm and 0.7cm,
    switch/.style={draw, rectangle, minimum width=1.8cm, minimum height=0.6cm, rounded corners, fill=blue!10, font=\scriptsize},
    host16/.style={draw, rectangle, minimum width=1.1cm, minimum height=0.45cm, rounded corners, fill=green!10, font=\tiny, align=center},
    host20/.style={draw, rectangle, minimum width=1.1cm, minimum height=0.45cm, rounded corners, fill=green!30, font=\tiny, align=center},
    hostw/.style={draw, rectangle, minimum width=1.1cm, minimum height=0.45cm, rounded corners, fill=green!10, font=\tiny, align=center},
    ap/.style={draw, rectangle, minimum width=1.3cm, minimum height=0.5cm, rounded corners, fill=yellow!10, font=\scriptsize},
    attacker/.style={draw, rectangle, minimum width=1.0cm, minimum height=0.45cm, rounded corners, fill=red!20, font=\scriptsize, thick},
    controller/.style={draw, rectangle, minimum width=1.9cm, minimum height=0.75cm, rounded corners, fill=purple!10, font=\scriptsize},
    csms/.style={draw, rectangle, minimum width=1.3cm, minimum height=0.45cm, rounded corners, fill=orange!10, font=\scriptsize},
    line/.style={-{Stealth[length=1.5mm]}, thin},
    controlLine/.style={dashed, -{Stealth[length=1.5mm]}, thin, purple!70}
]

\node[controller] (ryu) at (-7, 0) {Ryu Controller};
\node[font=\scriptsize\bfseries, above=0.15cm of ryu] {Control Plane};

\node[switch] (sw4) at (0, 0) {Switch 4 ($s_4$, Aggregate)};
\node[font=\scriptsize\bfseries, right=0.4cm of sw4] {Data Plane};

\node[switch] (sw1) at (-4.2, -2.2) {Switch 1 ($s_1$, Wired)};
\node[switch] (sw2) at ( 0, -2.2) {Switch 2 ($s_2$, WiFi)};
\node[switch] (sw3) at ( 4.2, -2.2) {Switch 3 ($s_3$, CSMS)};

\draw[line] (sw4) -- (sw1);
\draw[line] (sw4) -- (sw2);
\draw[line] (sw4) -- (sw3);

\node[hostw] (w1) at (-5.2, -3.8) {$C_1$};
\node[hostw] (w2) at (-4.2, -3.8) {$C_2$};
\node[hostw] (w3) at (-3.2, -3.8) {$C_3$};
\draw[line] (sw1) -- (w1);
\draw[line] (sw1) -- (w2);
\draw[line] (sw1) -- (w3);

\node[ap] (ap) at (0, -3.8) {WiFi AP};
\draw[line] (sw2) -- (ap);

\node[host16] (c1) at (-3.4, -5.2) {$C_4$};
\node[host16] (c2) at (-2.4, -5.2) {$C_5$};
\node[host16] (c3) at (-1.4, -5.2) {$C_6$};
\node[host16] (c4) at (-0.4, -5.2) {$C_7$};
\node[host16] (c5) at ( 0.6, -5.2) {$C_8$};

\node[host20] (c6)  at (-3.4, -6.3) {$C_9$};
\node[host20] (c7)  at (-2.4, -6.3) {$C_{10}$};
\node[host20] (c8)  at (-1.4, -6.3) {$C_{11}$};
\node[host20] (c9)  at (-0.4, -6.3) {$C_{12}$};
\node[host20] (c10) at ( 0.6, -6.3) {$C_{13}$};

\node[attacker] (atk) at (1.9, -5.2) {Attacker};

\draw[line] (ap) -- (c1);
\draw[line] (ap) -- (c2);
\draw[line] (ap) -- (c3);
\draw[line] (ap) -- (c4);
\draw[line] (ap) -- (c5);
\draw[line] (ap) -- (c6);
\draw[line] (ap) -- (c7);
\draw[line] (ap) -- (c8);
\draw[line] (ap) -- (c9);
\draw[line] (ap) -- (c10);
\draw[line] (ap) -- (atk);

\node[csms] (csms) at (4.2, -2.9) {CSMS};
\draw[line] (sw3) -- (csms);

\draw[controlLine] (ryu.east) -- node[above, font=\tiny, sloped] {OpenFlow} (sw4.west);
\draw[controlLine] (ryu.south) -- (sw1.north west);
\draw[controlLine] (ryu.south east) -- (sw2.north west);
\draw[controlLine] (ryu.east) -- (sw3.north west);

\node[draw, rounded corners, fill=gray!5, font=\tiny, align=left, inner sep=2pt]
    at (5.6, -3.8) {
    \begin{tabular}{ll}
    \tikz\draw[fill=green!10] (0,0) rectangle (0.2,0.14); & OCPP 1.6 Charger \\
    \tikz\draw[fill=green!30] (0,0) rectangle (0.2,0.14); & OCPP 2.0.1 Charger \\
    \tikz\draw[fill=orange!10] (0,0) rectangle (0.2,0.14); & CSMS Server \\
    \tikz\draw[fill=red!20]   (0,0) rectangle (0.2,0.14); & Attacker Host \\
    \tikz\draw[fill=blue!10]  (0,0) rectangle (0.2,0.14); & SDN Switch \\
    \tikz\draw[purple!70, dashed, thick] (0,0) -- (0.3,0); & OpenFlow Channel \\
    \end{tabular}
};

\end{tikzpicture}
}
\caption{SDN-enabled EVCI Topology. Switch
         $s_1$ hosts three wired OCPP~1.6 chargers ($c_1$--$c_3$);
         $s_2$ hosts ten WiFi chargers --- five running OCPP~1.6
         ($c_4$--$c_8$) and five running OCPP~2.0.1 Profile~2
         ($c_9$--$c_{13}$) --- connected via a WiFi AP, plus the
         co-located attacker; $s_3$ hosts the CSMS; and $s_4$ is
         the aggregate switch.}
\label{fig:topology_final}
\end{figure}

Figure~\ref{fig:topology_final} illustrates the EV charging
infrastructure used in this experiment.

The system comprises three component types: the controller, switches,
and hosts. The SDN controller handles rule installation, path
selection, and monitoring; we use Ryu, chosen for its low latency
relative to NOX, POX, Beacon, and
ONOS~\cite{holscher2022evaluation}. SDN switches forward traffic
under controller-defined rules; we use Open vSwitch (OVS) for
OpenFlow compatibility with Ryu. Let $SW$ be the set of all
switches: though functionally identical, we distinguish them (e.g.,
central switch, switch~2) by the hosts each connects to. The
topology consists of four OVS instances on a flat subnet, with
separation enforced via seven SDN policy rules rather than IP
subnetting. The experiment runs in
Mininet~\cite{lantz2010network}, with OVS configured to a
software-enforced TCAM limit of $N_r = 1{,}000$ flow entries,
approximating hardware switches that drop requests once the table
is full. Table~\ref{tab:openflow_flow} in
Appendix~\ref{app:tables} shows an example OpenFlow flow entry,
illustrating how the switch enforces the 60\,s idle timeout and
tracks flow statistics used to detect anomalies. Switch $sw2$ hosts the WiFi segment: ten chargers
($c_1$--$c_{10}$, five running OCPP~1.6 and five running
OCPP~2.0.1~P2) connected via a WiFi AP, plus ten paired EV emulators
for energy metering. Switch $sw1$ hosts three wired OCPP~1.6
chargers with three EV emulators; $sw3$ hosts the CSMS; $sw4$ is
the aggregate switch. Hosts belong to the Charging Point Operator (CPO), which operates
backend communication between the CSMS and the charging stations
$C$. The CSMS monitors charger status, schedules sessions, issues
start/stop commands, and defines session liveness with a timeout of
$T_{sess}=90$\,s: a charger missing this window is marked offline.
OCPP~1.6 remains the most widely deployed version; OCPP~2.0.1 adds
mandatory TLS security profiles (Profile~2: server-certificate
verification; Profile~3: mutual TLS). We use Profiles~2 and~3 to
resemble an up-to-date deployment. Each charger runs a real OCPP
implementation with inbound TCP servers on ports 80, 443, 22, and
8080.

\subsection{Attack method}

The attack proceeds in three phases: scanning, timeout exploration,
and low-rate filling and rejuvenation. Network parameters such as heartbeat period, idle timeout, table capacity, background
rate, are summarized in 
Appendix~\ref{app:tables}.

{\textbf{Phase 1: Scanning}}
The attacker joins the site WiFi using credentials obtained from a
public charging map or by brute-forcing the WPA3 passphrase,
receiving an IP via DHCP. Connected to the access point, the
attacker co-locates with the chargers and runs a TCP SYN scan across
the subnet on common charger ports. Switch $s_2$ ($Sw_2$ in
Figure~\ref{fig:topology_final}) hosts the access point and all WiFi
chargers, and SDN Policy R.1 permits East-West traffic within $s_2$,
so all WiFi chargers are reachable. The wired chargers on $s_1$ and
the CSMS on $s_3$ ($Sw_c$) are unreachable, and packets toward them
are silently dropped by Policy R.2. Each SYN installs a flow rule in
the switch. Port responses identify active chargers, and the
round-trip time is noted. The attacker sends probes with a fixed source port
against each identified charger(e.g., Scapy). The first probe may be handled by the
controller, introducing extra processing delay. Subsequent probes hit
the installed rule and bypass the controller, giving a clean measure
of raw network latency. The difference between the two reveals the
controller's processing overhead, which is used in Phase~2 to
distinguish a live rule from an expired one. These two reference
timings and the fixed source port are passed to Phase~2.

{\textbf{Phase 2: Timeout Exploration.}}
Using the reference timings from Phase~1, the attacker determines how
long the switch keeps a flow rule before discarding it. This idle
timeout governs how frequently the attacker must send keep-alive
traffic in Phase~3 to sustain the attack. The method relies on a simple observation: a probe that arrives while
a rule is still installed receives a fast reply from the switch
hardware; a probe that arrives after the rule has been discarded is
forwarded to the controller for reinstallation, producing a noticeably
slower reply. The attacker uses this timing difference as a binary
signal --- fast means the rule is alive, slow means it just expired. To locate the exact timeout, the attacker performs a binary search.
Each round begins by resetting the rule's idle timer, then waiting in
silence for a test duration, then sending a probe. A fast reply means
the rule survived the wait, so the search narrows upward; a slow reply
means the rule expired, so the search narrows downward. The process
repeats until the timeout value is pinpointed and carried into
Phase~3.

{\textbf{Phase 3: Low-Rate Filling and Rejuvenation}}
In Phase~3 the attacker uses the timeout learned in Phase~2 to
gradually fill the switch routing table, eventually preventing
legitimate charger traffic from being processed. Each new connection uses a different source port, which the switch
treats as a distinct flow and handles by installing new routing rules.
The attacker opens connections in small batches, sending a partial
request to hold each connection open, and refreshes each one with a
lightweight keep-alive message just before its rule would expire. A
small fraction of connections are deliberately dropped to blend in
with normal session turnover. As open connections accumulate, the routing table fills. Once it
reaches capacity the switch can no longer process new legitimate
flows, and the controller begins rejecting incoming charger traffic.
Algorithm~\ref{alg:phase3} in Appendix~\ref{app:tables} gives the
formal procedure, that is evaluated in Section~\ref{sec:eval}.

\section{Evaluation}
\label{sec:eval}
We structure the evaluation around four research questions that
together assess the impact of the threat and the effectiveness of
the proposed defense.

\begin{itemize}
    \item[\textbf{RQ1.}] Can a low-rate DoS attack negatively impact
    EV charging availability?
    \item[\textbf{RQ2.}] Does CS-SHIELD negatively impact normal
    charging operation?
    \item[\textbf{RQ3.}] To what extent does CS-SHIELD mitigate a
    low-rate DoS attack?
    \item[\textbf{RQ4.}] How well does CS-SHIELD protect EV charging availability under a low-rate DoS attack, and what overhead does each of its phases introduce?
\end{itemize}



\subsection{Evaluation Scenarios}
\label{subsec:scenarios}

We define four scenarios, summarised below, to isolate the effect
of the attack and of CS-SHIELD independently.

\begin{enumerate}

    \item \textbf{Scenario~1 --- Baseline (no attack, no MTD).}
          All ten WiFi chargers exchange OCPP messages such as
          Heartbeats, MeterValues, and StatusNotifications with the
          CSMS. No adversary is present and MTD is not active.

    \item \textbf{Scenario~2 --- Attack only (no MTD).}
          The attacker executes the three-phase Slow TCAM Exhaustion
          attack against $s_2$: \emph{Phase~1} TCP SYN scan, \emph{Phase~2}
          idle-timeout binary-search probe, \emph{Phase~3} low-rate
          Slowloris fill with rejuvenation. No detection or shuffling
          mechanism is active. We measure the
          attack's impact on availability, rule count, and
          packet forwarding delay.

    \item \textbf{Scenario~3 --- Attack with MTD.}
          The attacker executes the same  attack with the
          MTD detection and IP shuffling mechanism being active. The
          controller polls switches for flow statistics,
          cross-references against CSMS-reported charger IPs,
          detects the discrepancy, applies DROP, and triggers IP
          shuffling. We measure how effectively the
          defense contains the attack.

    \item \textbf{Scenario~4 --- MTD only (no attack).}
          MTD detection polling and IP shuffling run continuously
          under normal OCPP traffic. No attacker present. This
          scenario isolates the performance overhead introduced by
          the polling loop and shuffle trigger on legitimate
          sessions.

\end{enumerate}

\subsection{Metrics}
\label{subsec:metrics}

We will use the three metrics described below to evaluate our experiments.

\textbf{Flow-Table Rule Count} The total TCAM entry count on $s_2$ at time $t$ is

\begin{equation}
\small{     F(t) \;=\; F_{\mathit{leg}}(t) \;+\; F_{\mathit{atk}}(t)
    }\label{eq:flowcount}
\end{equation}

\noindent where $F_{\mathit{leg}}(t)$ counts legitimate charger rules
and $F_{\mathit{atk}}(t)$ counts attacker-injected rules, both
measured directly from the switch's flow table at time $t$. The
table saturates once $F(t) = N_r$, the TCAM capacity. The attacker's
objective is to drive $F_{\mathit{atk}}(t)$ up until $F(t)$ reaches
$N_r$, at which point the switch can no longer install new
legitimate rules.

\textbf{Availability} We consider OCPP sessions to be persistent WebSocket connections
that last for hours and must not be interrupted; availability is
therefore the property the attack directly targets, and the defense
must not reduce it further. A periodic shuffle would violate this
requirement, since changing addresses on a fixed schedule would tear
down active sessions at each interval and force reconnection,
indistinguishable from the attack's own effect. Site availability is the fraction of chargers with an active CSMS
session:

\begin{equation}
\small{    A(t) \;=\; \frac{1}{N_c}
    \sum_{i=1}^{N_c}
    \mathbf{1}\!\left[
        \text{charger } i \text{ has an active CSMS session at } t
    \right]
    }\label{eq:avail}
\end{equation}

\noindent A charger's CSMS session is considered active as long as
its flow-table rule has not expired and it continues to deliver
Heartbeats within $T_{sess}$. Because $F(t) \to N_r$ prevents new or
renewed legitimate rules from being installed, table saturation is
what drives $A(t) \to 0$; we therefore use $F(t)$ as a leading
indicator and $A(t)$ as the resulting outcome measure.

\textbf{Forwarding Delay} The one-way forwarding delay from charger $i$ to the CSMS is
estimated from round-trip probe timestamps:

\begin{equation}
\small{
    D_i(t) = \frac{T_{\mathit{rx},i}(t) - T_{\mathit{tx},i}(t)}{2}
    }\label{eq:delay}
\end{equation}

\noindent Mean forwarding delay across all $N_c$ chargers is:

\begin{equation}
\small{
    \bar{D}(t) = \frac{1}{N_c} \sum_{i=1}^{N_c} D_i(t)
    }\label{eq:meandelay}
\end{equation}

\noindent Any transient spike in $\bar{D}(t)$ caused by the purge
and re-injection step must remain below $T_{sess} = 90$\,s to ensure
OCPP session continuity.

\subsection{Results}
\label{sec:results}
\subsubsection{RQ1: Can Low-Rate DoS Negatively Impact EV Charging
Availability?}
\label{subsec:rq1}

\begin{figure}[htbp]
  \centering
  \begin{subfigure}[b]{0.49\columnwidth}
    \centering
    \includegraphics[width=\linewidth]{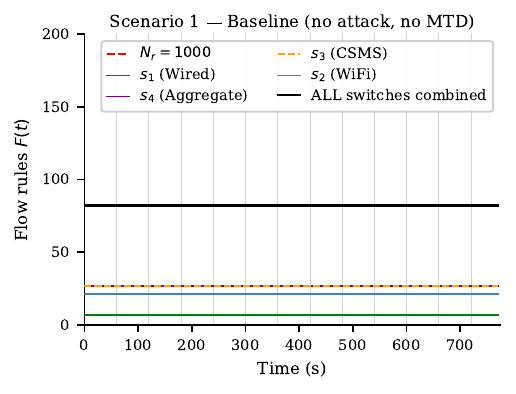}
    \caption{Sc.\,1 baseline.}
    \label{fig:rules_sc1}
  \end{subfigure}\hfill
  \begin{subfigure}[b]{0.49\columnwidth}
    \centering
    \includegraphics[width=\linewidth,height=12em]{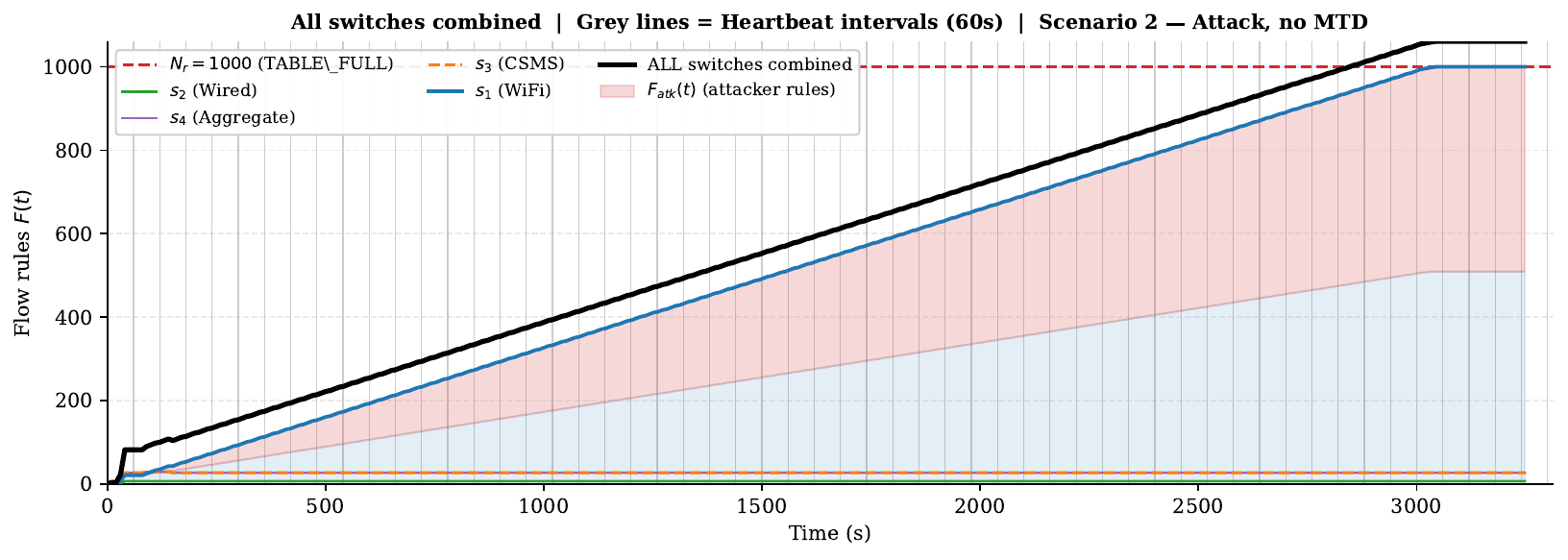}
    \caption{Sc.\,2 attack.}
    \label{fig:rules_sc2}
  \end{subfigure}\hfill
  
        \caption{Flow-table rules remain below capacity at
  baseline(a), rise to $N_r=1{,}000$ under attack(b)}
\end{figure}

Scenario~1 establishes the undisturbed baseline: after topology
startup, $F(t)$ stabilises across all switches and remains stable
throughout the experiment, well below $N_r = 1{,}000$, as shown
in Figure~\ref{fig:rules_sc1}. The WiFi switch $s_2$ --- the
attack target in Scenarios~2 and~3 --- carries the highest
individual rule count at 21 rules, reflecting its ten connected
chargers. The wired switch $s_1$ holds 7 rules for its three
chargers, while $s_3$ (CSMS) and $s_4$ (Aggregate) each hold
27 rules, giving a combined stable baseline of 82 rules across
all switches. 

Scenario~2 then evaluates the
attack with no defense active. During Phases~1 and~2 the attack is
entirely invisible --- $A(t)=1.0$ and the rule count shows no anomaly.
Phase~3 changes this: the attacker drives the rule count linearly to
full capacity at $t\approx3{,}350$\,s (Figure~\ref{fig:rules_sc2}),
after which the network can no longer accommodate new legitimate flows.
The full three-phase attack takes $\approx64$\,min total: Phase~1
nmap scan ($\approx13$\,min), Phase~2 binary-search timeout inference
($\approx1$\,min), and Phase~3 Slowloris fill with 245 connections
($\approx50$\,min).

\begin{figure}[htb]
    \centering
    \includegraphics[width=0.48\linewidth]{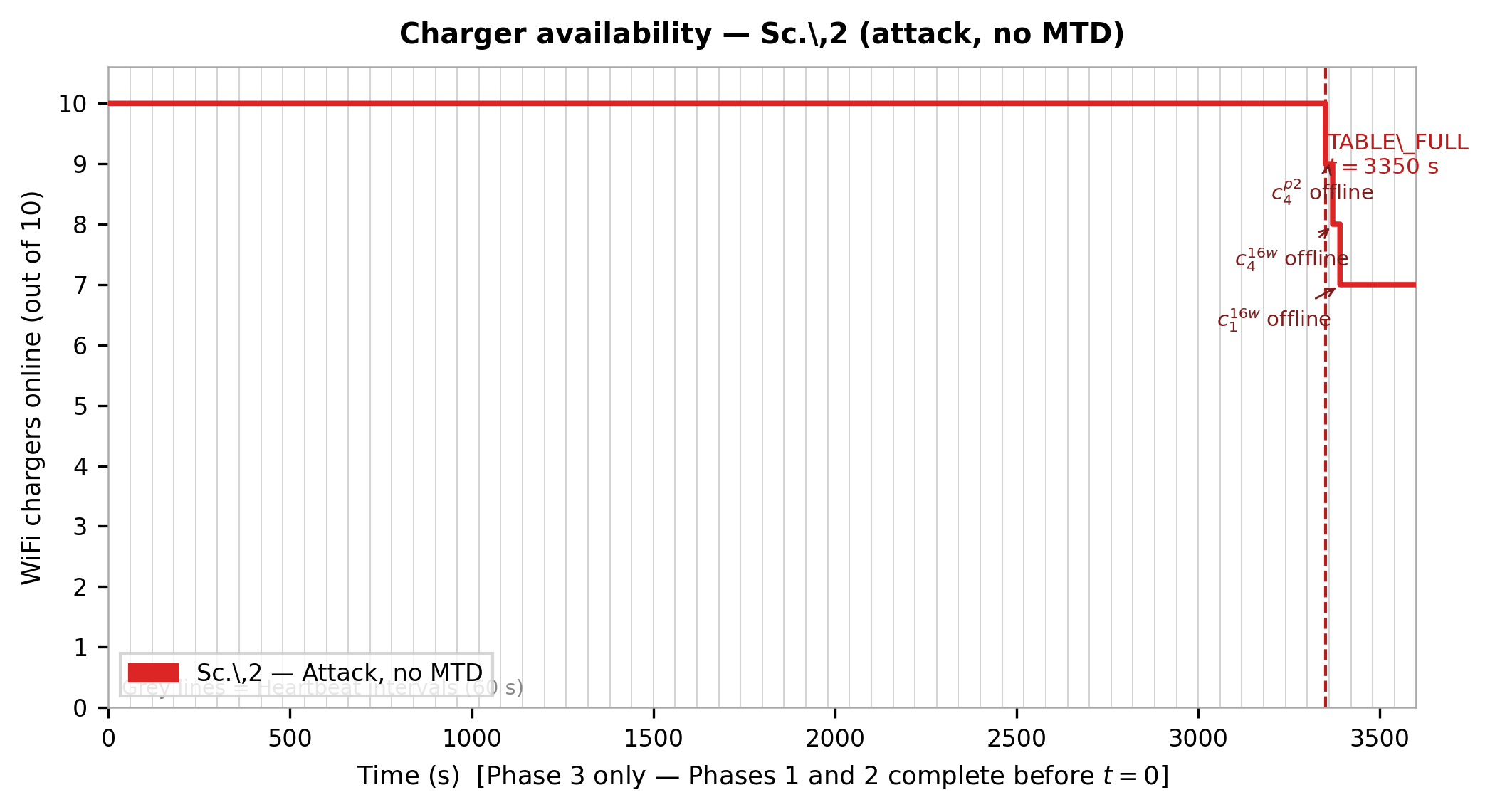}
   
      \caption{Saturation reduces
  charger availability to 7/10 within 90\,s.}
   \label{fig:availability_sc2}
  \end{figure}

Figure~\ref{fig:availability_sc2} shows that all ten chargers remain
fully operational throughout Phase~3 fill. The consequences arrive
suddenly at saturation: three idle chargers lose their CSMS sessions
in rapid succession as Heartbeat messages can no longer be delivered,
reducing site availability to 70\,\%. The remaining seven stay online
only because active EV sessions renew their routing entries frequently
enough to avoid expiry. Under low-utilisation conditions, such as
overnight when all chargers are idle, all ten would go offline within
90\,s of saturation. Recovery without intervention is impossible, as
the attacker continuously refills every freed slot. The answer to RQ1
is \textbf{yes} --- a low-rate DoS attack can silently disable a
charging site without generating any volume-detectable traffic.

\subsubsection{RQ2: Does CS-SHIELD Negatively Impact Normal Operation?}
\label{subsec:rq2}

\begin{figure}[htbp]
    \centering
    \begin{subfigure}[b]{0.49\columnwidth}
      \includegraphics[width=\linewidth]{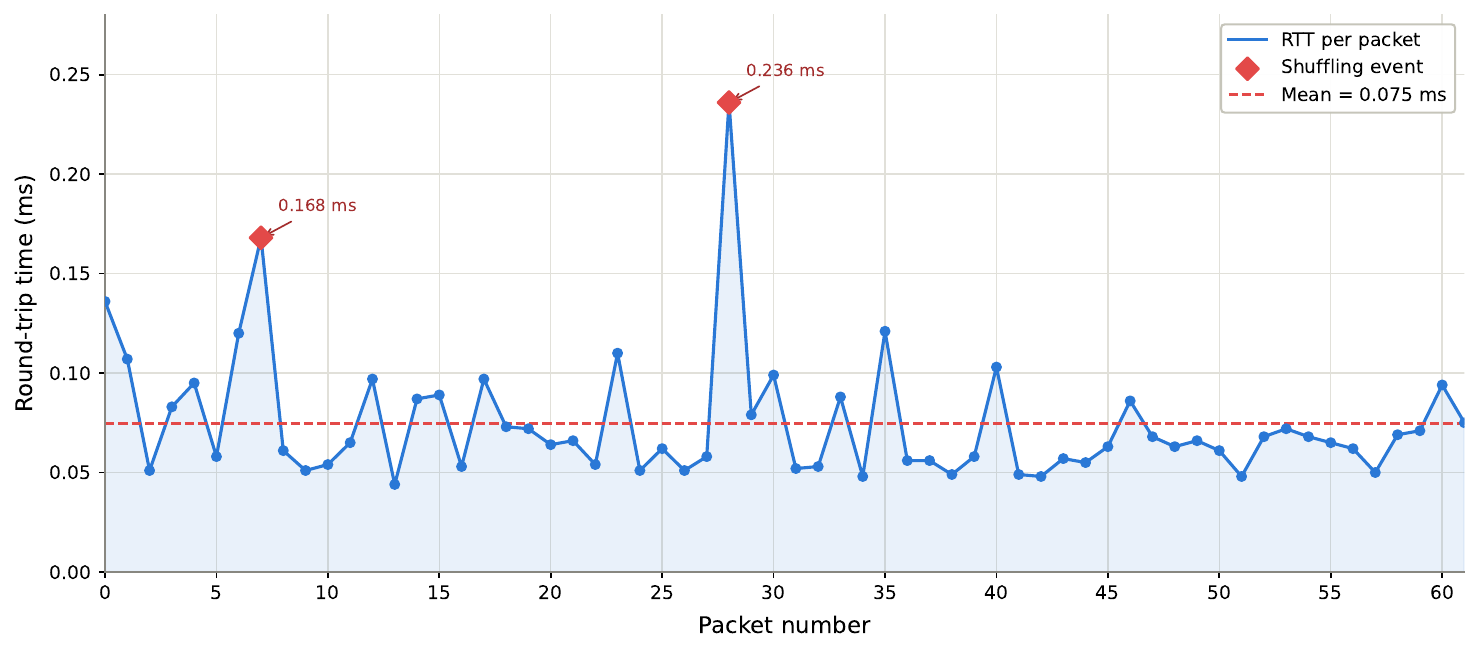}
      \caption{Charger-to-CSMS RTT.}
      \label{fig:rtt_sc4}
    \end{subfigure}\hfill
    \begin{subfigure}[b]{0.49\columnwidth}
      \includegraphics[width=\linewidth]{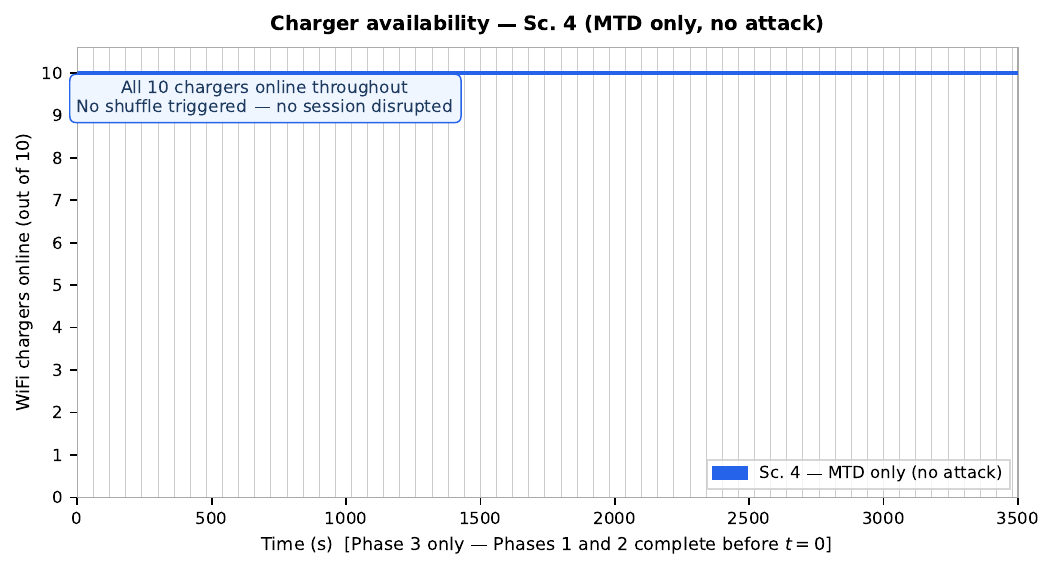}
      \caption{Charger availability $A(t)$.}
      \label{fig:avail_sc4}
    \end{subfigure}
    \caption{Scenario~4 (MTD active, no attacker): (a) RTT stays
             near a mean of 0.075\,ms, with the annotated peak
             coinciding with an IP shuffle event; (b) all ten WiFi
             chargers maintain $A(t)=1.0$ throughout, including
             across shuffle events.}
    \label{fig:sc4_combined}
\end{figure}

Scenario~4 evaluates CS-SHIELD with MTD enabled and no attacker
present. A continuous ICMP probe was issued from a charger to the CSMS, generating one packet per second over
the duration of the experiment. RTT measurements are presented in
Figure~\ref{fig:rtt_sc4}, and charger availability over time is shown
in Figure~\ref{fig:avail_sc4}.

The RTT series exhibits a brief elevation in the first two packets,
reaching approximately 0.14\,ms, attributable to reactive OpenFlow
flow installation on first contact rather than MTD overhead. Following
this initial transient, RTT stabilises into a consistent baseline
ranging from 0.044\,ms to 0.168\,ms, with a mean of 0.075\,ms. The IP shuffle event is visible at packet~28, where RTT reaches its
maximum observed value of 0.236\,ms --- a $3.1\times$ momentary
increase over the mean, corresponding to an absolute overhead of
0.161\,ms. RTT returns immediately to baseline on the next packet,
with no sustained elevation in the post-shuffle trace. On the availability dimension, the IP shuffle operates simultaneously
across all ten WiFi chargers connected to $s_2$. Despite the
purge and VIP remapping, all ten chargers maintain $A(t)=1.0$
throughout the experiment, as confirmed by Figure~\ref{fig:avail_sc4}.
Zero packet loss and zero charger disruptions were recorded.
Reconnection for any charger whose flow is evicted during the purge
occurs within the idle timeout period and thus does not break session
continuity, because the controller reinstalls the table-miss rule
immediately after the purge, allowing the reactive SDN path to resume
forwarding before the new VIP rules propagate. In the context of OCPP, where WebSocket sessions operate at heartbeat
intervals of 60--120\,s, a sub-millisecond single-packet RTT elevation
is imperceptible at the protocol level and carries no risk of session
termination. The MTD mechanism is therefore \emph{operationally
transparent} to legitimate traffic. The answer to RQ2 is \textbf{no}
--- CS-SHIELD does not negatively impact the normal operation of the
EV charging infrastructure.

\subsubsection{RQ3: To What Extent Does MTD Mitigate Low-Rate DoS?}
\label{subsec:rq3}

\begin{figure}[htbp]
  \centering
  \begin{subfigure}[b]{0.49\columnwidth}
    \centering
    \includegraphics[width=\linewidth]{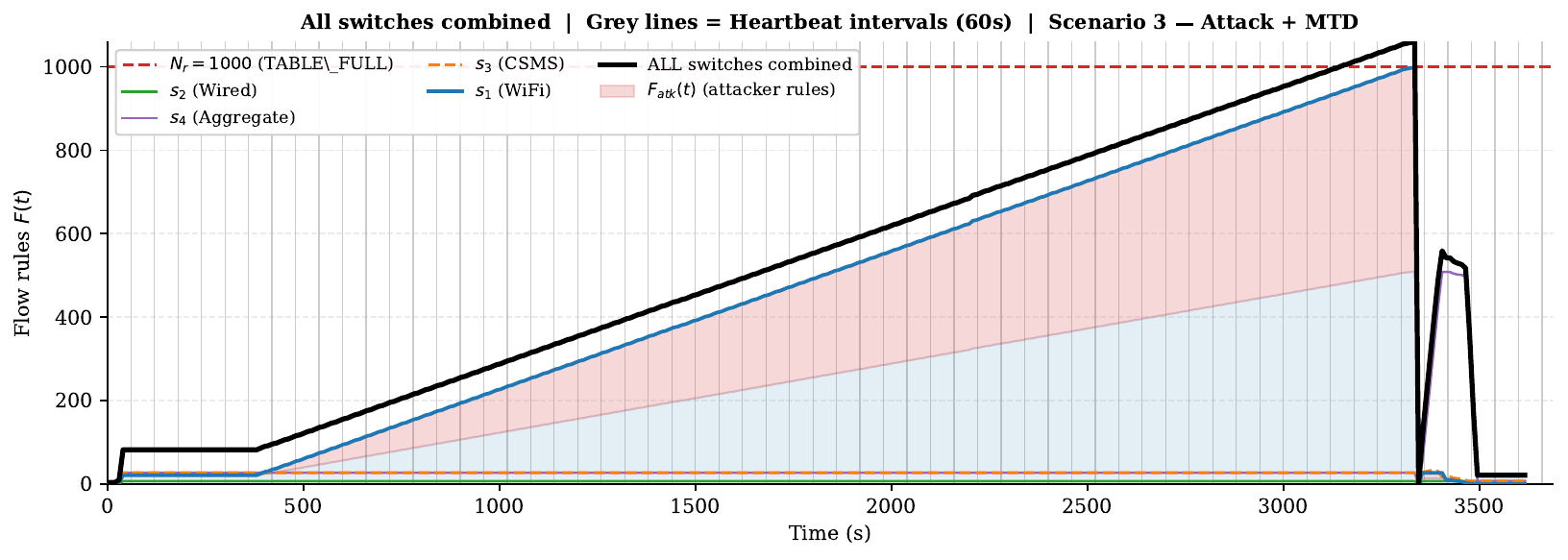}
    \caption{Flow-table rules.}
    \label{fig:rules_sc3}
  \end{subfigure}\hfill
  \begin{subfigure}[b]{0.49\columnwidth}
    \centering
    \includegraphics[width=\linewidth]{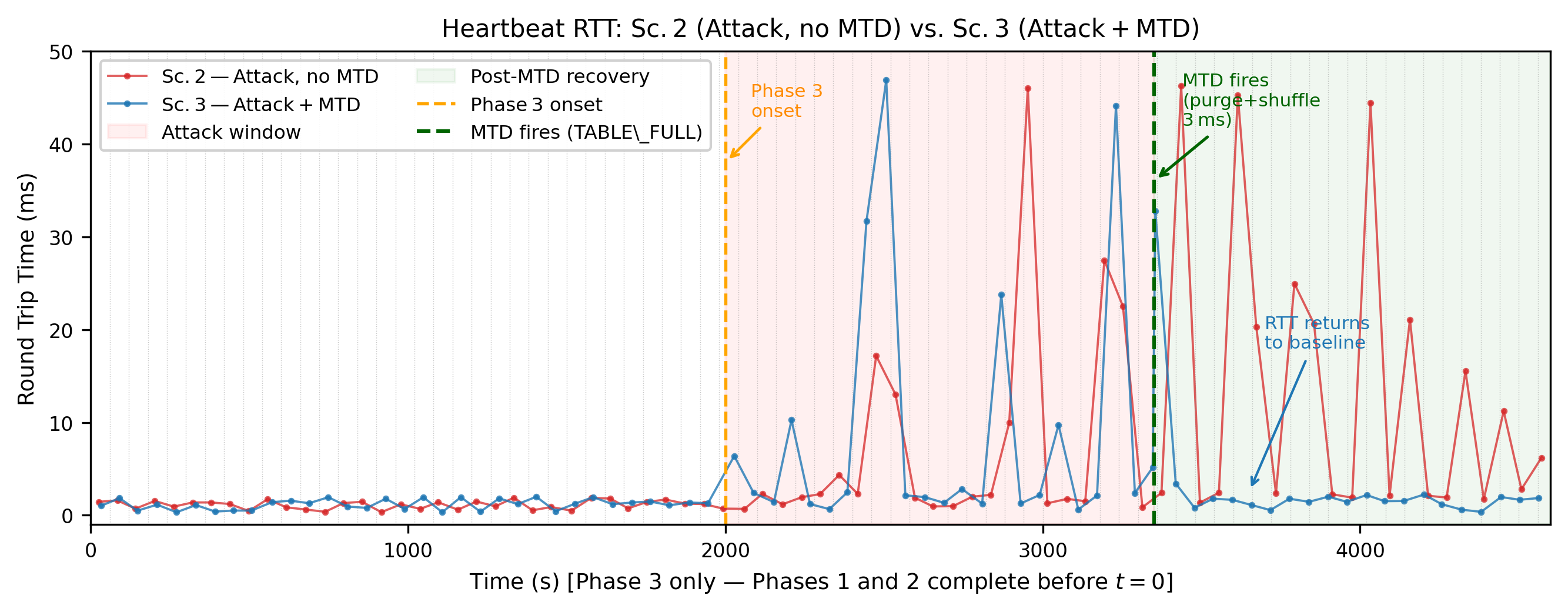}
    \caption{Heartbeat RTT.}
    \label{fig:rtt_comparison}
  \end{subfigure}
  \caption{Saturation triggers
  DETECT$\rightarrow$PURGE$\rightarrow$SHUFFLE; flow-table
  pressure is removed(a) and RTT returns to baseline(b).}
  \label{fig:rq3_combined}
\end{figure}

Scenario~3 evaluates CS-SHIELD with the attacker active.
Figure~\ref{fig:rq3_combined} presents the evolution of flow rules and
RTT across the full experiment duration. From $t=0$ to
$t\approx400$\,s, the rule count remains low and stable, reflecting
only legitimate OCPP sessions. At $t\approx400$\,s the attacker
begins injecting flows (pink region $F_{atk}(t)$), rising steadily
toward $N_r=1{,}000$ and mirroring Scenario~2 for $\approx3{,}000$\,s,
confirming no overhead pre-saturation. At $t\approx3{,}350$\,s the
TCAM reaches $N_r$, triggering DETECT$\to$PURGE$\to$SHUFFLE in
17.9\,ms: all attacker flows are dropped and virtual IPs reassigned to
all ten WiFi chargers, dropping the rule count to near zero in one
polling interval. The brief spike at $t\approx3{,}390$\,s is charger
re-registration under the new VIP mapping, completing within one
heartbeat interval; all subsequent attacker connections fail as its
real IP is blocked by the DETECT-phase DROP rule. RTT mirrors Scenario~2 until TABLE\_FULL, then returns immediately to
baseline (0.3--5\,ms), with a single transient spike of at most
42.9\,ms on re-registration. CS-SHIELD thus evicts all 
attacker-injected rules and restores normal forwarding within one
heartbeat interval. When the adversary is persistent, i.e., they try to restart the attack, after being thwarted, the shuffling mechanism in MTD forces them to repeat the attack from scratch. Without shuffling they would have reused the scanning results and proceed to DoS phase directly, or by initiating a new connection.

\subsubsection{RQ4: How Well Does MTD Protect EV Charging Availability
When Subject to Low-Rate DoS Attack?}
\label{subsec:rq4}


While shuffling, the CSMS re-registers chargers under their new VIP mappings within one
heartbeat interval, before the 90\,s session timeout. Thus, no charger
is counted as unavailable: $A(t)=1.0$ throughout the attack,
response, and re-registration period. The charger re-registration completes within one heartbeat interval, therefore it does not count as a session loss. The CSMS path
is restored before the session timeout is reached, so no charger is
counted as unavailable. CS-SHIELD therefore provides complete availability protection under the evaluated low-rate attack.  To measure the overhead of each part of our mechanism, we repeated the experiments in three cases: \emph{detect-only},
\emph{purge-only} (DROP plus table clearing), and the full
CS-SHIELD process. DETECT costs
0.265\,ms mean. PURGE costs 2.37\,ms in the purge-only variant and
1.48\,ms in the full variant. The difference is due to host scheduling jitter in Mininet. SHUFFLE adds 16.4\,ms, for a total of
17.9\,ms. This experiment was repeated 7 times. Full results can be found in the Appendix~\ref{app:tables}.  

\subsection{Discussion}
\label{subsec:discussion}

After MTD fires, the attacker's entire $\approx64$\,min of
reconnaissance is invalidated in 17.9\,ms: every IP binding learned
during Phases~1-2 becomes worthless once each charger gets a new
address drawn randomly from a pool of over 65{,}000 candidates.
Re-discovering all addresses at the same low rate takes far longer
than the interval between purges, so the attack cannot succeed again
without fundamentally changing its approach. The partial impact in Scenario~2, where only three of ten chargers
went offline, reflects the heterogeneous nature of charging sessions:
chargers with active EV sessions renew their flows frequently and
resist saturation, while only idle chargers relying solely on the
60\,s Heartbeat are immediately vulnerable. In a sustained attack, all
chargers would eventually be affected as sessions complete. 

There are limitations to our approach. Real hardware differs in three respects: (1) TCAM rules incur
different hardware costs depending on how many packet header fields
they match (our single-field IP matches are among the least
expensive); (2) hardware eviction policies may differ from OVS
idle-timeout expiry; and (3) OpenFlow latency on hardware is
typically lower than in Mininet.

CS-SHIELD shuffles only when the flow table is exhausted, at which
point the attack has already put the affected sessions at risk. A
time-periodic shuffle would instead tear down active sessions at
fixed intervals and force reconnection, indistinguishable from the
attack's own effect.
\section{Conclusion}
\label{sec:conclusion}
This paper presents CS-SHIELD, a Moving Target Defense mechanism for
SDN-enabled EV charging infrastructure that defends against low-rate
DoS attacks by cross-referencing the switch's flow state against the
CSMS's authenticated session list, detecting the discrepancy, and
responding with an IP address shuffle that evicts attacker flows and
invalidates prior reconnaissance. Experiments on an emulated SDN testbed with a real OCPP
implementation confirm: a low-rate DoS attack can silently reduce
site availability to 70\,\% within 90\,s of flow-table saturation
with no prospect of recovery (RQ1); CS-SHIELD adds only
sub-millisecond RTT overhead and maintains $A(t)=1.0$ under normal
operation (RQ2); it detects and mitigates the attack at saturation,
restoring normal forwarding within one heartbeat interval (RQ3); and
availability remains at $A(t)=1.0$ fleet-wide throughout the attack,
contrasting with the 70\,\% floor observed without defense (RQ4).
The defender--attacker asymmetry is the central outcome: over an
hour of reconnaissance is invalidated in milliseconds, and
rebuilding it takes far longer. A proactive flow-arrival-rate
detector could shorten the pre-saturation window; validation on
hardware SDN switches and formal quantification of shuffle entropy
are left for future study.

\appendix
\section{Testbed Symbols, Example Flow Entry, and Attack Algorithm}
\label{app:tables}

Table~\ref{tab:def_in} defines the symbols used to model the system
components: the switches ($Sw_2$, $Sw_c$, $Sw_R$), the sets of nodes
and chargers ($N$, $C$), and the control and access entities ($K$,
$AP$, $W$, CSMS).

\begin{table}[h]
\centering
\caption{System Element Definitions}
\label{tab:def_in}
\scriptsize
\setlength{\tabcolsep}{3pt}
\begin{tabular}{cl}
\toprule
\textbf{Symbol} & \textbf{Definition} \\
\midrule
$Sw_2$                  & Dedicated switch connected to Access Point \\
$Sw_i\ (i \in \mathbb{N}^+)$ & Other switches (excluding $Sw_2$) \\
$Sw_c$                  & Central switch connected to CSMS \\
$Sw_R$                  & $\{Sw_i \mid i \in \mathbb{N}\}$: set of other switches \\
$N$                     & $\{n_j \mid j \in \mathbb{N}\}$: set of all nodes \\
$C$                     & $\{c_i \mid i \in \mathbb{N}\}$: set of all chargers \\
$K$                     & SDN controller \\
$AP$                    & WiFi Access Point, connected only to $Sw_2$ \\
$W$                     & $\{w_k \in N \mid k \in \mathbb{N}\}$: hosts connected to Access Point \\
\bottomrule
\end{tabular}
\end{table}

Table~\ref{tab:openflow_flow} shows an example OpenFlow flow entry.
This flow instructs the switch to forward packets from port 1 to
port 2 with the specified source and destination MAC addresses. The
timeout enforces rule expiration: 60 seconds of inactivity removes
the entry. The counters track flow duration, packet count, and byte
volume, enabling the controller to monitor active flows and detect
anomalies.

\begin{table}[H]
\centering
\caption{Example OpenFlow Flow Entry}
\label{tab:openflow_flow}
\scriptsize
\renewcommand{\arraystretch}{0.85}
\setlength{\tabcolsep}{2pt}
\begin{tabular}{ll}
\toprule
\textbf{Field} & \textbf{Value} \\
\midrule
Cookie / Timeout / Priority & \texttt{0x0} / \texttt{60\,s} / \texttt{1} \\
\midrule
In port / Src MAC & \texttt{1} / \texttt{00:00:00:00:00:01} \\
Src port / Dest MAC & \texttt{1000} / \texttt{00:00:00:00:00:02} \\
Dest port & \texttt{80} \\
\midrule
Action & \texttt{output:2} \\
Duration & \texttt{764.169\,s} \\
Packets / Bytes & \texttt{242{,}538} / \texttt{13{,}851{,}836{,}796} \\
\bottomrule
\end{tabular}
\end{table}

Table~\ref{tab:params} summarizes the symbols and parameters used in
the attack model and its evaluation.

\begin{table}[t]
\centering
\caption{Network parameters}
\label{tab:params}
\setlength{\tabcolsep}{3pt}
\renewcommand{\arraystretch}{1.0}
\scriptsize
\begin{tabularx}{\linewidth}{p{0.4cm}Xr}
\toprule
\textbf{Sym.} & \textbf{Description}\\
\midrule
$T_{hb}$ & OCPP Heartbeat period.\\
$T_o$    & Idle timeout; set equal to $T_{hb}$.\\
$N_r$    & Max flow rules on $s_1$.\\
$N_c$    & WiFi chargers on $s_1$.\\
$W_b$    & Background rate $N_c/T_{hb}$; attacker upper bound.\\
$D_w$    & $T_{hb}/N_c$; avg.\ gap between Heartbeats. \\
$D_{lt}$ & Safety margin; $\Delta_{SL}=T_o-D_{lt}$. \\
$F_{max}$& $\lfloor N_c(T_o-D_{lt})/T_{hb}\rfloor$; max sustained flows. \\
$T_{sat}$& $\lceil N_r/(2B)\rceil\cdot(B/W_b)$ at $B=F_{max}$. \\
\bottomrule
\end{tabularx}
\end{table}

Algorithm~\ref{alg:phase3} gives the formal Phase~3 low-rate filling
and rejuvenation procedure summarized in
Section~\ref{sec:setup}.

\begin{algorithm}[t]
\scriptsize
\caption{Phase 3: Low-rate filling and rejuvenation}
\label{alg:phase3}
\KwIn{$d$, $\hat{T}_o{=}60$\,s, $\Delta_{SL}{=}58$\,s, $B{=}9$,
$\epsilon{=}0.1$, $W_b{=}0.167$\,p/s, $s_{start}{=}5000$}
\KwOut{$R_{atk}(t)\to N_r$; $A(t)\to 0$}
$k\leftarrow 0$; $\mathcal{S}\leftarrow\emptyset$\;
\While{\textbf{active}}{
  $t_0\leftarrow now()$; $T_{round}\leftarrow B/W_b$\;
  \For{$j=1$ \KwTo $B$}{
    $s_k\leftarrow s_{start}+k$; $k\mathrel{+}=1$\;
    Handshake$(d,80,s_k)$; send partial GET\;
    $\mathcal{S}\leftarrow\mathcal{S}\cup\{(s_k,\,t_0+\Delta_{SL})\}$\;
  }
  $due\leftarrow\{(s,t)\in\mathcal{S}:t\leq t_0+\Delta_{SL}\}$
  sorted by $t$\;
  \ForEach{$(s_j,t_j)$ in first
  $\lceil(1{-}\epsilon)|due|\rceil$ of $due$}{
    Send \texttt{X-a:b\textbackslash r\textbackslash n} on
    $\mathrm{conn}_{s_j}$; $t_j\leftarrow now()+\Delta_{SL}$\;
  }
  \ForEach{remaining in $due$}{
    Close socket;
    $\mathcal{S}\leftarrow\mathcal{S}\setminus\{(s_j,t_j)\}$\;
  }
  Sleep until $t_0+T_{round}$\;
}
\end{algorithm}

\begin{figure}[H]
  \centering
  \begin{subfigure}[b]{0.32\columnwidth}
    \centering
    \includegraphics[width=\linewidth]{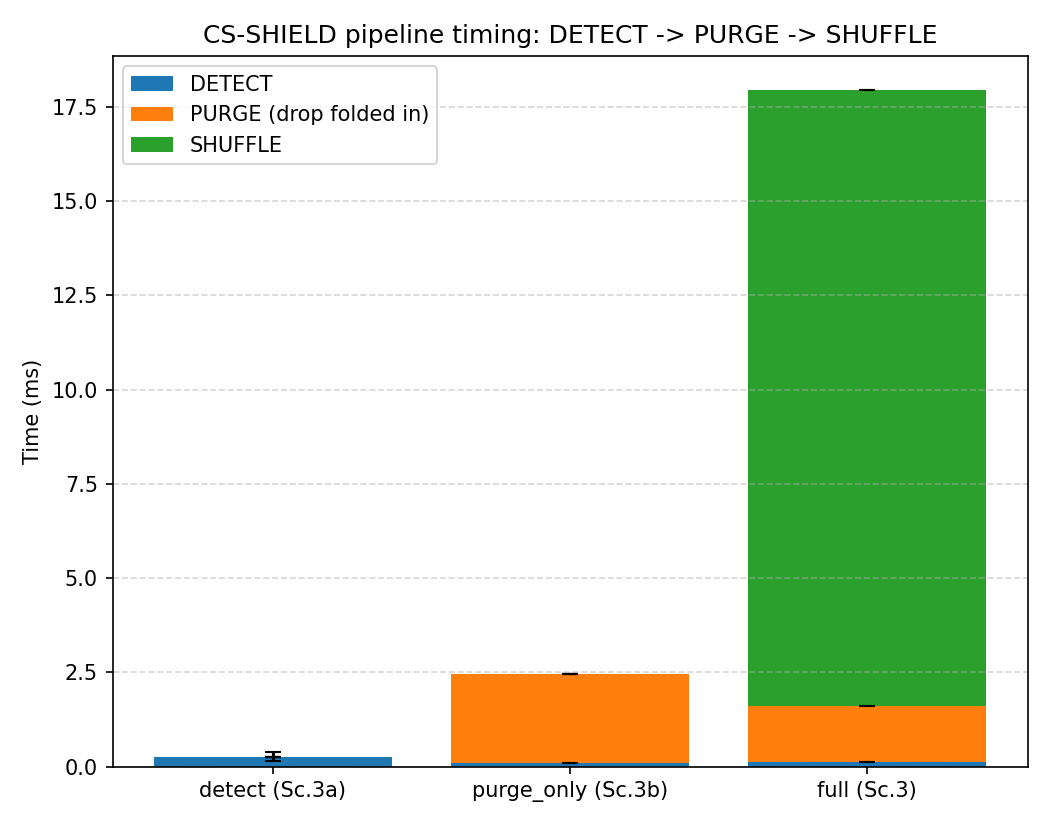}
    \caption{Enforcement cost.}
  \end{subfigure}\hfill
  \caption{SHUFFLE adds 16.4\,ms
  beyond PURGE for a total response of 17.9\,ms}
  \label{fig:ablation_combined}
\end{figure}

\paragraph{\textbf{Ablation Study:}}
\label{app:ablation}
As explained in Results section, overhead cost of each phase is measured in three variants:\emph{detect-only},
\emph{purge-only}, and the full
CS-SHIELD workflow. Figure \ref{fig:ablation_combined} DROP blocks the currently observed attacker source but leaves the target address unchanged, so a new source can resume Phase~3 without
reconnaissance. SHUFFLE reassigns every charger address, invalidating
prior target knowledge and forcing the attacker to repeat scanning
and timeout inference. At the measured low rate this changes the
reconnaissance/recovery economics from roughly 50\,min per re-armed
attempt to approximately 4.6 days for rediscovery.

For polling overhead, Scenario~3 and Scenario~4 have mean poll-cycle
costs of $18.8\pm12.9$\,ms  and $20.4\pm13.7$\,
respectively. The overlapping variance indicates no measurable
additional polling burden during active defense.

\bibliographystyle{splncs04}

\bibliography{references}

\end{document}